# Quantum cascade laser-pumped terahertz molecular lasers: frequency noise and phase-locking using a 1560nm frequency comb

**Jean-Francois Lampin[1], Antoine Pagies[1], Giorgio Santarelli[2], Jeffrey Hesler[3], Wolfgang Hänsel[4], Ronald Holzwarth[4] and Stefano Barbieri[1*]**

[1]Institut d'Électronique, de Microélectronique et de Nanotechnologie, CNRS, Université de Lille, 59652 Villeneuve d'Ascq, France
[2] LP2N, IOGS, CNRS, Université de Bordeaux, 33400 Talence, France
[3] Virginia Diodes Inc., Charlottesville, USA
[4] Menlo Systems GmbH, Martinsried, Germany
* e-mail: stefano.barbieri@univ-lille.fr

**The recent demonstration of a terahertz (THz) molecular gas laser pumped by a mid-infrared quantum cascade laser (QCL) has opened up new perspectives for this family of sources, traditionally relying on $CO_2$-laser pumping. A so far open question concerning QCL-pumped THz molecular lasers (MLs) is related to their spectral purity. Indeed, assessing their frequency/phase noise is crucial for a number of applications potentially exploiting these sources as local oscillators. Here this question is addressed by reporting the measurement of the frequency noise power spectral density (PSD) of a THz ML pumped by a 10.3µm-wavelength QCL, and emitting 1mW at 1.1THz in continuous wave. This is achieved by beating the ML frequency with the 1080$^{th}$ harmonic of the repetition rate of a 1560nm frequency comb. We find a frequency noise PSD < 10Hz$^2$/Hz (-95dBc/Hz) at 100kHz from the carrier. To demonstrate the effect of the stability of the pump laser on the spectral purity of the THz emission we also measure the frequency noise PSD of a $CO_2$-laser-pumped 2.5THz ML, reaching 0.1Hz$^2$/Hz (-105dBc/Hz) at 40kHz from the carrier, limited by the frequency noise of the frequency comb harmonic. Finally, we show that it is possible to actively phase-lock the QCL-pumped molecular laser to the frequency comb repetition rate harmonic by controlling the QCL current, demonstrating a sub-Hz linewidth.**

# 1. Introduction

Recently, a novel approach for THz MLs was demonstrated, where conventional $CO_2$-laser pumping is replaced by a mid-infrared (MIR) QCL [1]. This results in a much more compact, frequency agile and deployable source that has revived the interest for THz MLs, in particular for their exploitation as low-phase noise sources operating at room temperature in applications such as radio-astronomy, atmospheric science, free-space communications, high-resolution spectroscopy or radar imagery [2-8].

Compared to $CO_2$–lasers, QCLs offer an unrivalled frequency agility that allows accessing an unprecedented number of molecular transitions. Indeed, these sources cover a very large spectra bandwidth, from ~3µm to ~11µm, with single-mode operation at room temperature, output powers in the 100mW range, and continuous tuning over several hundreds of GHz. This removes virtually any restriction on the choice of the molecular transitions that can potentially be used as gain medium, in particular allowing the exploitation of molecules with large dipole moments that are most of the times out of reach of $CO_2$-lasers, emitting only on discrete lines. For the laser exploited in this work, the use of a QCL has allowed to pump a transition of $NH_3$, a light molecule, characterized by a high permanent dipole moment, and a fast non-radiative relaxation [1]. Thanks to this, an output power of 1mW at 1.073THz has now been attained (see next Section and Fig.1(c)).

Another fundamental difference between $CO_2$-lasers and QCLs that must be considered when pumping a ML, is related to their linewidth, or frequency noise. Typical $CO_2$-laser linewidths are in the ~1-100kHz range, whereas MIR QCLs linewidths are approximately 100 times larger (~1-10MHz) [9-11]. Determining how this affects the spectral purity of QCL-pumped MLs is the main purpose of this work. To this end we have setup two experiments, both exploiting the harmonic of the repetition rate of a 1560nm FC as local oscillator (LO) to generate a beat-note (BN) in the radio-frequency (RF) range with (i) a 1.073THz, QCL-pumped, ML [1] and (ii) a commercial, $CO_2$-pumped, $CH_3OH$ ML emitting at 2.523THz. From these BNs we derive the corresponding frequency noise PSD spectra. Finally, we demonstrate that by controlling the QCL current, the QCL-pumped ML can be actively phase-locked to the FC harmonic, yielding a sub-Hz linewidth. These results pave the way to the development of QCL-pumped MLs as compact, low-phase noise THz sources operating a room temperature.

# 2. THz molecular laser

The schematic of the THz ML used in this work is described in Fig.1(b). A distributed feedback (DFB) QCL (AdTech Optics) emitting at 10.3µm wavelength (29THz) is focused into a cylindrical waveguide resonator, of 6mm diameter, and pumps the gain medium consisting of pure $NH_3$ maintained at a pressure of ~1-2 Pa (the ZnSe window on the right-hand side is used to seal the waveguide). In Fig.1(a) we report a simplified energy diagram, showing the relevant levels. The QCL is tuned on the *sa*Q(3,3) transition at 29THz (967.346 $cm^{-1}$- which is not reachable by a CW $CO_2$ laser), and lasing action at $\nu_{THz}$ = 1.073THz takes place between the *a*(3,3) and *s*(3,3) levels. These correspond to the antisymmetric and symmetric states of the so-called "umbrella-mode" transition, associated to the tunneling of the N atom through the potential barrier formed by the H atoms (respectively blue and white spheres in the inset of Fig.1(a)). Both the MIR and THz transitions present high dipole moments of 0.2D and 1.05D (the $NH_3$ molecule has a 1.42D static dipole moment). Moreover, the ammonia molecule has a high rotational constant and a fast relaxation rate, and was identified as an ideal molecule for THz lasing in the 1980s, without, however, the possibility to pump it with a CW $CO_2$ laser. [12-14]. As shown in Fig.1(c) all these factors allow reaching an exceptionally high MIR→THz conversion efficiency $\eta_{opt}$ = 1% (i.e. a tenfold increase compared to what was achieved in Ref.[1]). The ideal theoretical limit (lossless cavity, rapid relaxation, 100% absorption of pump photons, etc), is given by the expression $\eta_{opt} = {}^{1}/_{2}\left({}^{\upsilon_{THz}}/_{\upsilon_{MIR}}\right) = 1.8\%$ [15]. Our result corresponds to ~55% of this ideal limit and represents a twofold improvement compared the highest reported conversion efficiency for a $CO_2$-pumped ML laser (33% of theoretical limit at 2.5THz) [16].

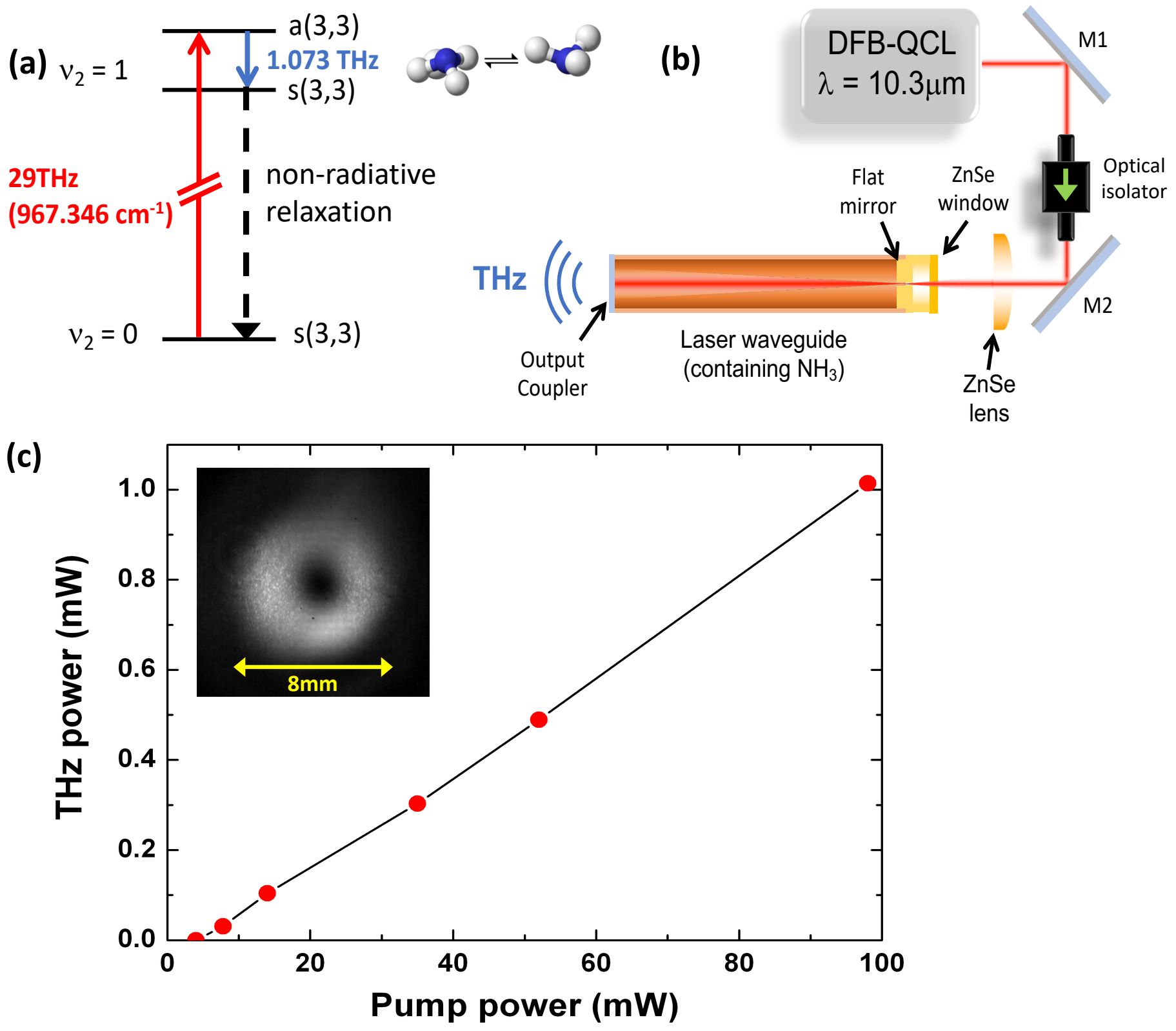

**Figure 1.** a) Energy diagram of the levels relevant for laser operation. b) Schematic of the QCL-pumped ML. The laser cavity is closed at one end by an output coupler, obtained by depositing a metallic grid on a Silicon substrate, and on the other end by a plane metallic mirror, with a 1.2mm hole drilled in its center. An isolator (~30dB isolation, and ~60% transmission) is used to reduce the optical feedback on the DFB QCL. c) THz ouput power vs MIR pump power, measured without the optical isolator displayed in panel (b). Inset: intensity plot of the optical beam collected with a microbolometer camera positioned at about 4cm from the laser output coupler.

Lasing takes place on the $TE_{01}$ waveguide mode, presenting an azimuthal symmetry, and a minimum on the waveguide axis [17]. This is indeed the mode with the lowest propagation losses since the electric field is perpendicular to the waveguide radius, and must therefore be zero at the metal surface to satisfy the boundary condition. In the inset of Fig.1(c) we report the intensity profile of the optical beam, measured with a THz microbolometer camera positioned at about 4cm from the laser output coupler, and without focusing optics. As shown in the Inset, the mode diameter is ~8mm, i.e. 2mm larger than the waveguide diameter, corresponding to a beam divergence of approximately 3deg.

## 3. Experimental setup

The experimental setup for the measurement of the frequency noise PSD of the 1.073THz ML is displayed in Fig.2. The purpose of this setup is to generate a beat-note (BN) between the frequency of the THz laser and the harmonic of the repetition rate of an optical frequency comb (FC). The latter is based on a custom made, harmonically mode-locked, fs-fiber laser emitting at 1560nm, with a repetition rate $f_{rep}$ = 1GHz, characterized by a very low level of frequency (phase) noise (see Supporting Information, Section S1 and Fig. S3). As a result, the frequency noise of the generated BN is dominated by the contribution from the ML, which can be therefore measured using a suitable frequency discriminator.

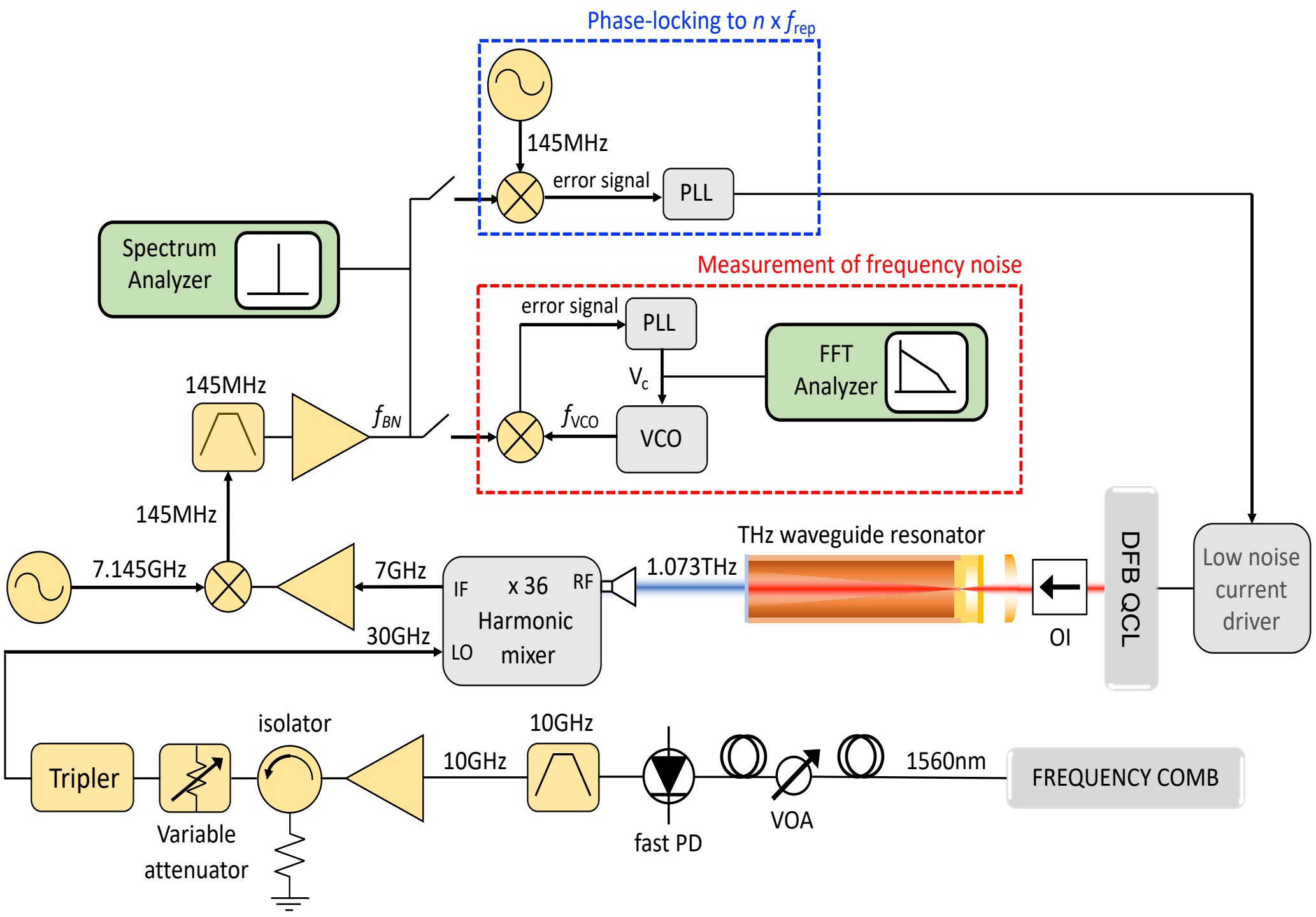

**Figure 2.** Schematic of the experimental setup used for the measurement of the frequency-noise PSD and of the phase-locking of the QCL-pumped ML (see main text). The beam from the FC is first attenuated through a variable attenuator (VOA) and then focused on a fast photodiode (PD). The down-converted BN at 145MHz is finally used to measure the frequency noise PSD of the ML (red frame) or for its phase-locking to the FC (blue frame).

The 10.3µm DFB QCL shown in Fig.2 is maintained at a temperature of 19.5C using a Peltier cooler and a temperature controller with mK precision, and is driven at ~1A (10V bias voltage) by a low noise current driver (Koheron, DRV110) with a current noise of 300pA/Hz$^{1/2}$. An optical isolator (OI) is placed right after the QCL to minimize optical feedback (see also Fig.1(b)). This is a crucial step in order to avoid destabilizing the QCL emission frequency that must be kept at the center of the ~100MHz-wide, Doppler-broadened $NH_3$ transition at 29THz (see Fig.1(a)). Inside the THz waveguide, we obtain a MIR pump power of ~60mW, yielding ~0.6mW at $\nu_{THz}$ =1.073THz at the output of the waveguide resonator. The THz beam is coupled, through a metallic rectangular waveguide terminated by a diagonal horn, to the RF input of a 750GHz-1100GHz sub-harmonic mixer (Virginia Diodes, VDI-WM250SAX), characterized by a SSB, RF to Intermediate Frequency (IF) total conversion loss of ~17dB at 1THz (~30dB intrinsic mixer conversion loss), and by an IF bandwidth of about 40GHz. The mixer horn is positioned at ~4cm from the ML output coupler without any optics. Given the symmetry of the azimuthal $TE_{01}$ laser beam (see Inset of Fig.1(c)), it was possible to couple to the $TE_{10}$ rectangular guided mode only a fraction of the total THz power. From the input area of the horn (1.6x1.6mm$^2$) and the measured mode intensity profile, we estimate a coupling coefficient of roughly 10%.

To generate the LO signal for the harmonic mixer we begin by focusing the beam of the FC on a fast photodiode (Discovery Semiconductors, DSC-50S) and then use a narrow band filter to select the 10$^{th}$ harmonic of the photocurrent signal at 10GHz = 10 x $f_{rep}$. The LO signal at 30GHz (+10dBm) is generated using a passive frequency tripler. The total LO multiplication factor of the WM250SAX being 36, the resulting BN produced at the mixer IF output has a frequency $f_{BN} = (36 \times 30) f_{rep} - \nu_{THz}$ = 1080– 1073 = 7GHz. $f_{BN}$ is finally down-converted to 145MHz using an RF mixer and a synthesizer.

## 4. Beatnote measurements

An example of BN power spectrum recorded with a spectrum analyzer (Fig. 2) in a single-shot is reported in Fig.3(a). We find an instantaneous linewidth of 10kHz, limited by the resolution bandwidth (RBW). From the acquisition sweep rate of 78MHz/s, such linewidth is recorded with an integration time of 0.13ms. Instead, on a time scale of several 100ms, the BN displays a slow drift of several 100kHz that we attribute to the combined effect of temperature/mechanical fluctuations and low-frequency noise of the QCL current driver. The parasitic lines on both sides of the beatnote are harmonics of ~180kHz whose origin could not be identified.

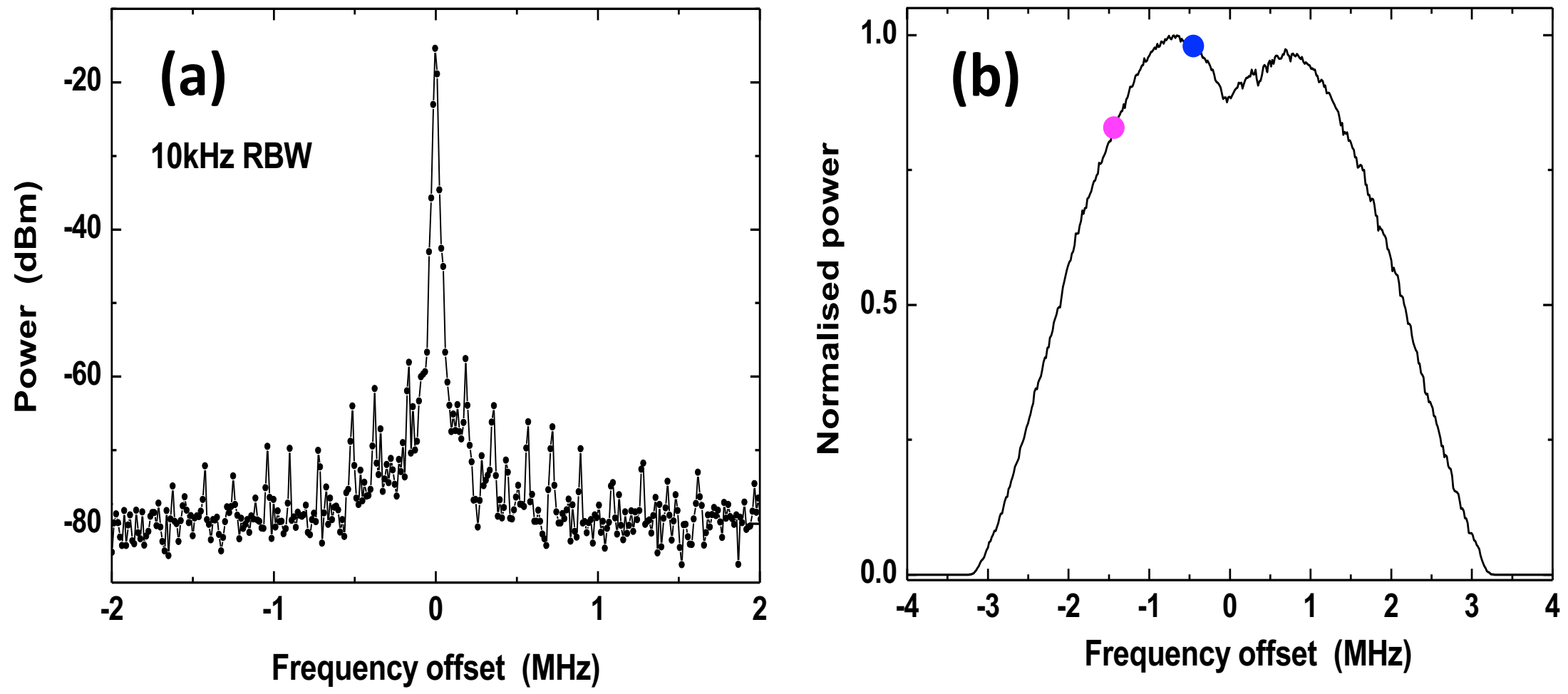

**Figure 3.** Beatnote spectra of the free-running QCL-pumped ML. The spectra were recorded with the Spectrum Analyser shown in Fig.2, and are offset by ~145MHz. a) Single shot spectrum recorded with a RBW of 10kHz and a sweep rate of 78MHz/s. b) BN spectrum recorded with the spectrum analyzer in Max Hold mode, and a RBW of 100kHz (see main text). We attribute the slight asymmetry of the spectrum to the fact that the cavity Fabry-Perot mode was not perfectly centered on the molecular transition. The blue and pink dots indicate approximately the points of minimum and maximum MIR → THz MT, and correspond approximately to the points where the frequency noise PSD spectra of the QCL-pumped ML were recorded (blue and pink traces in Fig.4).

In Fig.3(b) we report another BN spectrum that was recorded with the spectrum analyzer in Max Hold mode, and a RBW of 100kHz. To understand the meaning of this spectrum we note that at a pressure of ~1-2 Pa, the *sa*Q(3,3) pump transition is inhomogeneously broadened by the Doppler effect, with a FWHM of ~90MHz, i.e. larger than the QCL linewidth, that we estimate to <10MHz [18]. Therefore the QCL pumps only a class of molecules with a well-defined speed component along the axis of the THz waveguide resonator. This has two main consequences: (i) to a first approximation the THz gain has a bandwidth of a few 100kHz, given by the QCL linewidth divided by the ratio between the MIR and THz frequencies ($\frac{\upsilon_{MIR}}{\upsilon_{THz}}$ = 29.0THz/1.073THz = 27); (ii) changing the QCL frequency automatically induces a change of the lasing frequency, since the latter follows the peak of the THz gain curve [19]. This is precisely what we did to obtain the spectrum of Fig.3(b). In particular we first adjusted the length of the ML cavity using a piezo-actuator fixed to the output coupler (not shown in Fig.1(b)) in order to place a Fabry-Perot cavity mode as close as possible to the center of the THz molecular transition (from the reflectivities of the output coupler and of the input mirror of ~0.3, and ~0.95 respectively – see Fig.1(a) –, and neglecting waveguide losses, we obtain a FWHM ~ 60MHz for the Fabry-Perot mode). Then we swept the frequency of the QCL pump by changing its temperature by a few mK using the temperature controller. As shown in Fig.3(b), this generates a peak with a FWHM of ~4MHz, dominated by the Doppler broadening of the THz transition [19].

Another significant feature in the spectrum of Fig.3(b) is the ~10% drop of THz power at the center of the broad peak, that we attribute to the well-known Lamb dip transfer [14, 20-22]. This process is triggered by the fact that compared to pumping away from the center of the MIR transition, when the QCL is pumping right at its center, the *sa*Q(3,3) transition is saturated twice as hard by the two counterpropagating beams inside the cavity (the pump is not fully absorbed in one pass), producing a drop in the effective pumping rate which, in turns, produces a drop of THz power [18]. A necessary condition for the observation of a transferred Lamb dip in a ML is that the Doppler broadened pump transition is at least partially saturated. Given the divergence of the QCL beam inside the THz waveguide cavity (Fig.1(b)), a precise evaluation of the saturation intensity in our laser is difficult, however, from the data presented in Ref.[23] we estimate a saturation power in the ~1mW range for a collimated beam of a few mm waist, indicating that the saQ(3,3) transition should indeed be well saturated (see Supporting Information Section S2). A well-established consequence of the Lamb dip transfer is a modification of the real part of the complex THz susceptibility, inducing, through frequency pulling, a non-linear dependence of $\nu_{THz}$ on $\nu_{MIR}$ close to the center of the molecular transition. This, in turns, can produce important deviations of the modulation transfer MT $= \frac{\Delta\upsilon_{THz}}{\Delta\upsilon_{MIR}}$, from its value imposed by the Doppler effect, and given by $\frac{\upsilon_{THz}}{\upsilon_{MIR}} = 1/27$ [19]. The dependence of the MT on $\nu_{MIR}$ is related to a two-photon or Raman process [12, 24], and can be derived using density matrix equations applied to a 3-level system interacting with two counterpropagating waves [25, 26]. Such analysis was confirmed experimentally [25, 27], and predicts in particular two points of vanishing MT, symmetrically positioned with respect to the center of the molecular transition, and approximately located inside the dip just before the maxima of THz power (blue dot in Fig.3(b)).

## 5. Frequency-noise PSD and phase locking to the frequency comb

To measure the frequency noise PSD of the ML we used a demodulation technique based on a voltage-controlled oscillator (VCO – Minicircuit JCOS-175LN) [28]. As shown in the red-dashed frame of Fig.2, the VCO frequency, $f_{VCO}$, is directly phase-locked to $f_{BN}$ through a high-bandwidth phase-locked loop (PLL) circuit. As a result the VCO control voltage $V_C$, with a measured slope $\delta f_{VCO}/\delta V_C$ = 4.2MHz/V, is proportional to the sum of the frequency noises of $f_{VCO}$ and $f_{BN}$. If the former (see Fig.4, purple trace) is negligible compared to the latter, the frequency noise PSD of $f_{BN}$ (= 1080 x $f_{rep}$ – $\nu_{THz}$) can be finally obtained by measuring the power spectrum of $V_c$ using a fast-Fourier transform analyzer (FFT in Fig. 1) and multiplying it by the VCO slope [4]. The frequency noise PSD of $f_{BN}$ will then correspond to that of the ML if the noise of the 1080$^{th}$ harmonic of $f_{rep}$ can be neglected.

In Fig.4 we report two representative frequency noise PSD spectra of the ML in the 100Hz-6MHz range (pink and blue traces). These were obtained by temperature tuning the QCL pump frequency so that during the acquisition the lasing frequency could be kept in two specific regions relative to the center of the molecular transition. Their position is indicated approximately by the blue and pink dots on the left of the line center (negative offset) in Fig. 3(b): operating the ML between these two points yielded values of frequency noise between the pink and blue traces. We attribute this finding to the Lamb dip induced non-linear dispersion described in the previous Section. In particular, when operating in correspondence to the blue and pink dots, the MT is respectively minimized and maximized, resulting into a lower and higher frequency noise [25, 27].

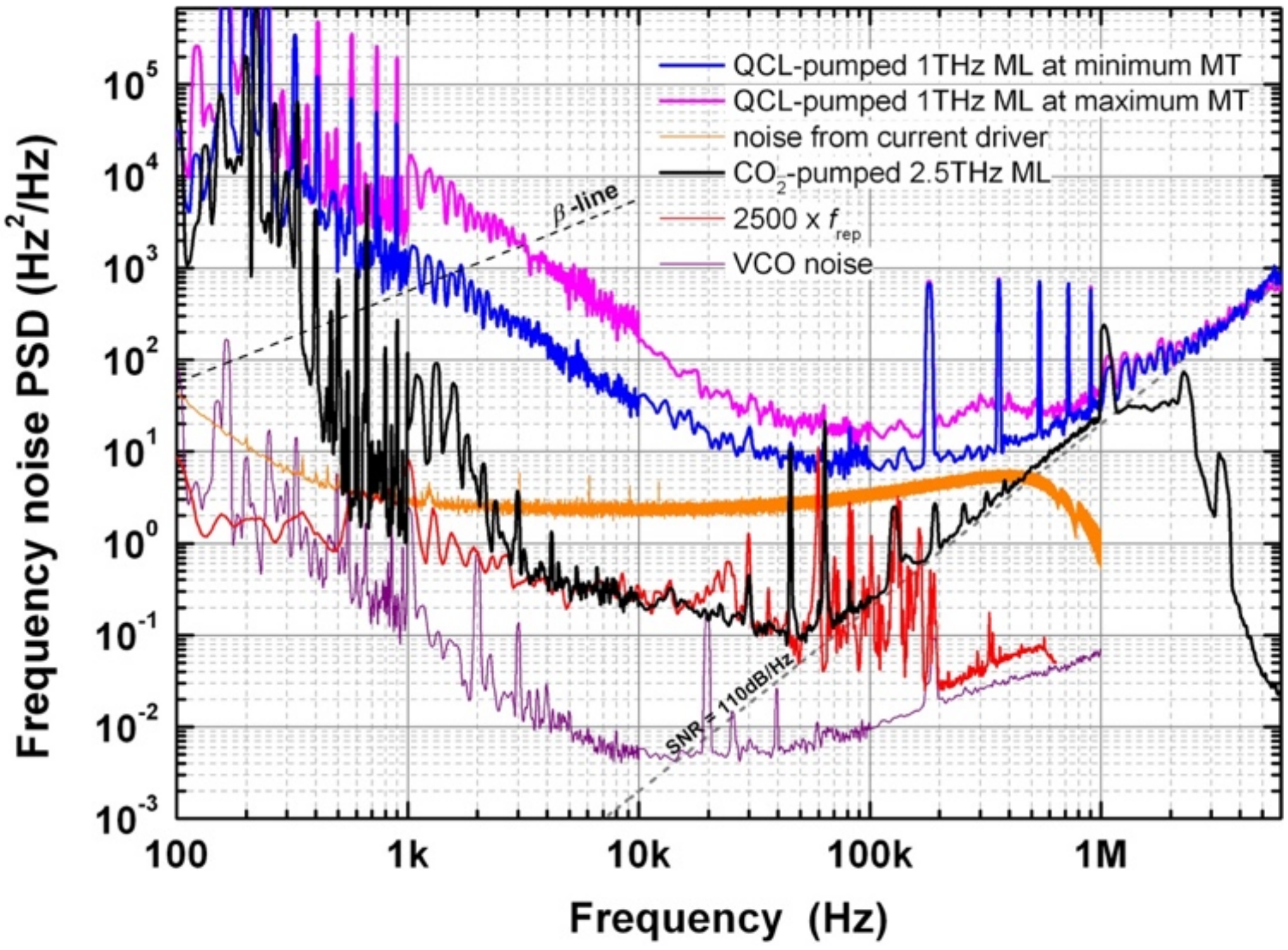

**Figure 4.** Frequency noise PSD spectra. The blue and pink traces are the frequency noise PSDs of the QCL-pumped, 1.073THz ML, measured with the setup of Fig.2. They were recorded in correspondence to the blue and pink dots in Fig.3(b). The orange spectrum is the estimated noise contribution from the QCL current driver assuming a MT = $\frac{\upsilon_{THz}}{\upsilon_{MIR}}$. The black trace is the frequency noise PSD of the 2.5THz line of a commercial $CO_2$-pumped methanol laser. The red spectrum corresponds to the frequency noise PSD of 2500 x $f_{rep}$ (See Supporting Information, Section S1 and Fig. S1 and S2). The purple spectrum is the frequency noise PSD of the VCO shown in Fig.2.

As expected, both frequency-noise PSD spectra present an increase at low frequency. Above ~20kHz, we observe a flattening at ~ 7Hz$^2$/Hz (blue spectrum) and 17Hz$^2$/Hz (pink spectrum). Finally, beyond ~1MHz, both spectra reach asymptotically the $f^2$ growth determined by the signal to noise ratio (SNR) of 110dB/Hz. The orange trace represents the estimated noise contribution from the QCL current driver, obtained by multiplying the measured driver current noise by the product of the static current tuning coefficient of the QCL (~180GHz/A) times the MT. Here we assumed a MT of $(27)^{-1}$, i.e. equal to $\frac{\upsilon_{THz}}{\upsilon_{MIR}}$ , which, as regards to the blue trace, should be an over-estimate (see the previous Section). Considering also the fact that the current tuning coefficient is expected to drop from *dc* to 100kHz [29], overall we expect the orange curve to represent an upper limit for the current driver noise contribution. Despite these considerations, we cannot exclude the possibility that the current driver noise is responsible for the observed flattening of the frequency noise PSD spectra above a few tens of kHz. On the other hand, we can certainly rule out that the latter is determined by the Shawlow-Townes limit of the QCL, which is in the ~100Hz$^2$/Hz (yielding ~0.1Hz$^2$/Hz at 1THz) and could be observed only in the work of Bartalini et al. at Fourier frequencies >10MHz [30]. At low frequencies, the frequency-noise PSD of MIR QCLs follows typically a $1/f$ trend. A survey of the published literature on QCLs operating at wavelengths between 4.6μm and 10.3μm at room temperature [9, 10, 29, 31, 32] provides rather scattered values for the frequency noise PSD, in the range $10^5$-$10^7$Hz$^2$/Hz and $10^4$-$10^6$Hz$^2$/Hz, respectively at 1kHz and 10kHz from carrier. Assuming a MT = $\frac{\upsilon_{THz}}{\upsilon_{MIR}}$ these numbers are compatible with the values of ~3x10$^3$Hz$^2$/Hz at 1kHz and ~100Hz$^2$/Hz at 10kHz, obtained from the average of the measured blue and pink spectra in Fig.4. Finally, by integrating the frequency noise PSD of the ML up to the crossing with the $\beta$-separation line (~1.3kHz) [33], we obtain, for an integration time of 10ms, linewidths of 23kHz and 80kHz, respectively for the blue and pink traces in Fig.4. We note that these linewidths, limited by technical noise, are presently dominated by the spurs below 1kHz and would be reduced by at least a factor ~4 without.

To further quantify the effect of the pump source on the ML frequency noise we have also measured the frequency noise PSD of a commercial, $CO_2$-pumped methanol ML (Edimburgh Instruments) emitting at 2.523THz. In this case, to generate the beatnote between the THz frequency and the FC repetition rate harmonic we have used an electro-optic detection technique instead of an electronics mixer (see Refs. [28, 34]). A description of the technique and a schematic of the experimental setup are provided in Supporting Information, Section S3 and Fig. S4. To derive the frequency-noise PSD from the beat-note signal we have employed the same demodulation technique used for the ML: the resulting spectrum is shown in Fig.4 (solid black trace). We can clearly identify three regions. Up to ~3kHz we observe the technical noise of the FIR laser, while above ~50kHz we hit the SNR of our detection. From the $\beta$-separation line crossing (~350Hz), we obtain a linewidth of 8kHz for an integration time of 10ms. In the range ~3kHz-100kHz the noise analysis is instead limited by the frequency noise of the harmonic of the FC repetition rate (~ 2500 x $f_{rep}$). This noise contribution is represented by the solid red trace in Fig.4. The latter was obtained by (i) beating the FC with a fiber laser at $\lambda$ = 1542nm (195THz) stabilized to a high finesse cavity, and (ii) by multiplying the corresponding FNPSD by the factor $(2.5/195)^2$, to take into account the reduction of frequency noise from 195000 x $f_{rep}$ to 2500 x $f_{rep}$ (see Supporting Information, Section S1 and Fig. S1 and S2 for more details). At 40kHz we observe a frequency noise of 0.1Hz$^2$/Hz, i.e. more than 20dB below that achievable at 2.5THz by frequency multiplication of a state-of-the-art 10GHz commercial generator (R&S®SMA100B-B711N, see Fig.S5) [35]. To the best of our knowledge this represents the lowest reported value for a THz ML [27, 36]. Based on the frequency noise of the QCL-pumped ML (10Hz$^2$/Hz @10÷100kHz Fourier frequencies) and on the fact that $CO_2$ lasers linewidthts are ~100 times below those of QCLs, the actual frequency noise of the $CO_2$-pumped methanol ML is expected as low as $10^{-3}$Hz$^2$/Hz (@10÷100kHz). Such value could be confirmed if the FC was locked to an optical reference, such as in Ref. [37], where the reached microwave stability corresponds to a frequency noise below $5\cdot10^{-5}$ Hz$^2$/Hz when scaled to 2.5 THz.

The frequency noise of the QCL-pumped ML can be actively reduced by phase-locking to the harmonic of $f_{rep}$. The schematic of the setup used is shown in Fig.2 (blue frame), and is obtained by phase-comparing $f_{BN}$ to the signal at 145MHz from an RF generator. The error signal is sent to a home-made PLL circuit used to control a fraction of the QCL current. In Fig.5 we report the power spectra of the locked beatnote recorded with RBWs of 10kHz (Fig.5(a)) and 1Hz (Fig.5(b)). From the noise bumps in the first spectrum we deduce a control bandwidth of 600kHz, not limited by the PLL electronics (the parasitic lines close to the carrier are the same found in Fig. 3(a) and Fig.4). The linewidth of the spectrum in Fig.5(b) is limited by the RBW, showing that actual linewidth is < 1Hz. By integrating the normalized power spectrum of Fig.5(a) from 1Hz up to 1MHz and by dividing by the RBW, we obtain a residual *rms* phase noise $\sigma_\varphi \sim 0.04$rad, corresponding to a fractional power of 99.8% of the BN signal coherently locked [31].

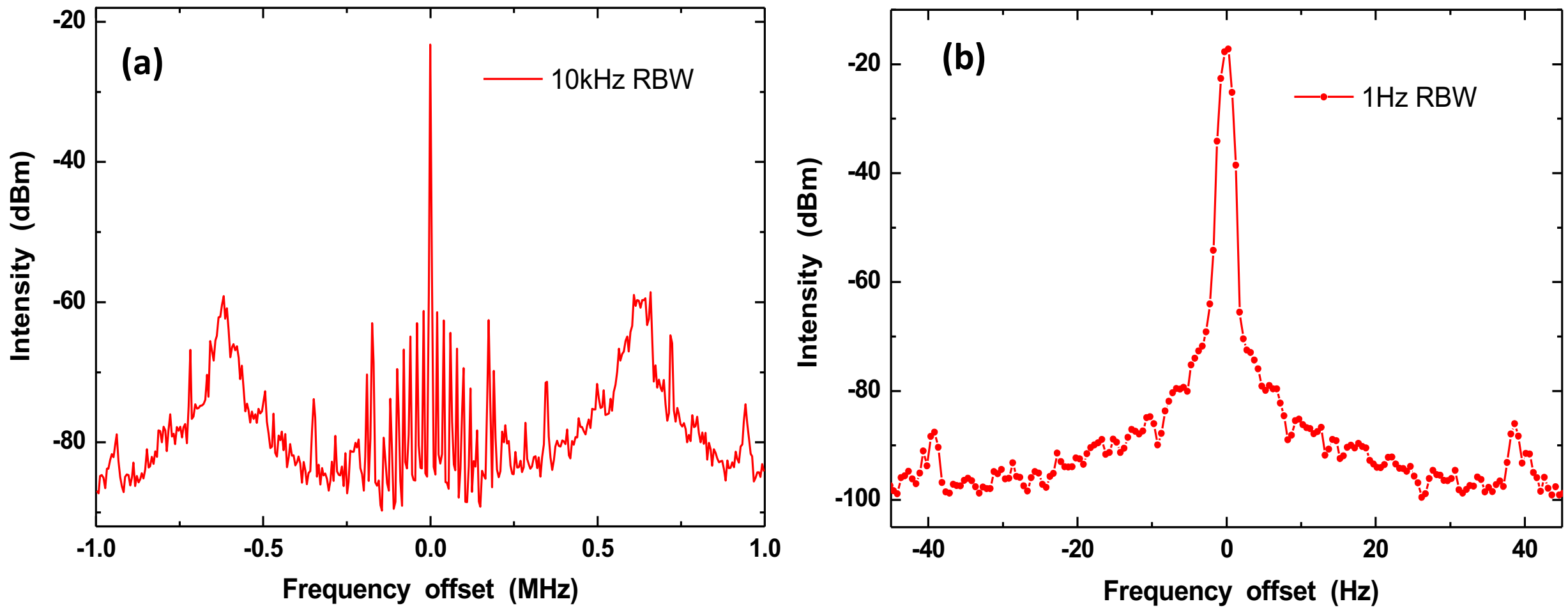

**Figure 5.** Power spectra of the BN frequency with the QCL-pumped ML phase-locked to the 1080th harmonic of $f_{rep}$. The spectra were measured with the Spectrum Analyzer of Fig.2. a) RBW = 10kHz. The noise bumps on the both sides of the central line indicate a servo bandwidth of ~600kHz. b) RBW = 1Hz.

## 6. Conclusion

QCL-pumped MLs are still at their early infancy, and have a bright future as compact THz sources operating at room temperature. On one hand, based on what has been achieved with $CO_2$-pumped MLs, extending their operation at least up to 7THz should not present major difficulties. On the other hand, by leveraging on the frequency agility of MIR QCLs, allowing the selection of the most appropriate molecules, it will be possible, as demonstrated by the performance of the laser studied in this work, to considerably surpass the MIR → THz conversion efficiencies of $CO_2$-pumped MLs, virtually guaranteeing the access to power levels in the mW range in CW. With the laser studied in this work, we reach a wall plug efficiency of $10^{-4}$ in CW at 1THz, which is comparable to what is achieved with state-of-the art frequency multiplication chains and photomixers (with photomixers the demonstrated generated power at 1THz is limited 10-20 μW), and approximately a factor of ten higher than what can be reached at room temperature with intra-cavity difference-frequency generation in MIR QCLs [38-41]. Finally, the fact that QCLs provide a continuous coverage of the entire MIR spectrum, partially mitigates the intrinsically poor tunability of THz MLs by facilitating the access to molecular transitions close to the targeted THz spectral region.

In this work we have addressed a so-far open question related to the spectral purity of these novel sources, in particular on how the latter is impacted by the use of a semiconductor laser pump. This is a fundamental issue for the type of applications that, we believe, are particularly adapted to QCL-pumped MLs, such as high-resolution spectroscopy, radio-astronomy, or, possibly, free-space THz communications [2-8, 42]. The results of our measurements show that despite QCLs being intrinsically noisier that $CO_2$ lasers, the frequency noise of QCL-pumped, THz MLs in free running remains extremely-low, below what has so far been achieved with cryogenically-cooled THz QCLs (although at higher frequencies, see Supporting Information, Section S4 and Fig.S5) [28, 43] and, for Fourier frequencies >100kHz, comparable to what could be obtained with multiplied THz sources pumped by state-of-the-art synthesizers (see Supporting Information, Section S4 and Fig. S5). We also demonstrated that, should these values not be enough, QCL's pumped MLs can be phase-locked to more stable oscillators rather straightforwardly by controlling the QCL current, compared to $CO_2$-pumped MLs, where phase- or frequency-locking requires a feedback on the cavity-length [27,44-46]. In these experiments, the harmonic of the repetition rate of a custom-made, 1560nm, harmonically mode-locked FC has been used as reference oscillator. By comparison with a $CO_2$-pumped 2.5THz ML, it was shown that the frequency noise PSD of this reference is $10^{-1}Hz^2/Hz$ (-105dBc/Hz SSB phase noise) at 40kHz from the carrier, which places an upper limit to the frequency noise of the ML and represents, to the best of your knowledge, the lowest reported for this type of sources [27, 36]. It is an important corollary of this work, that a free-running fiber-based FC can have a repetition rate with a sufficiently low phase-noise to serve as high-quality reference for THz metrology applications [47]. Even more, the noise level could be reduced to below $5{\cdot}10^{-5}$ $Hz^2/Hz$ if the FC was locked to an optical reference [37], offering yet a lower noise floor for frequency measurements in the THz domain.

## Acknowledgements

We gratefully acknowledge Benoît Darquié, Daniele Rovera, and Francois Rohart for helpful discussions. We acknowledge Jean Minet for performing the measurement of the current noise of the QCL current driver, and Ghaya Baili for the loan of the fast photodiode. This work was partially supported by the LABEX Cluster of Excellence FIRST-TF (ANR-10-LABX-48-01), within the Program "Investissements d'Avenir" operated by the French National Research Agency (ANR). We also acknowledge financial support from the Nord-Pas de Calais Regional Council and 'Fonds Européens de Développement Régional' (FEDER) through the 'Contrat de Projets Etat Région (CPER)' "Photonics for Society". This work has benefited from the facilities of the ExCELSiOR Nanoscience Characterization Center and the RENATECH network.

## Authors contributions

J-F.L. planned and performed the experiments on the QCL-pumped molecular laser, designed and built the laser and analyzed the data. A.P. built and optimized the QCL-pumped molecular laser. G.S. participated to the experiments, made the phase-lock electronics, and analyzed the data. J.H. gave assistance on the use of the sub-harmonic mixer and participated to the experiment on the QCL-pumped molecular laser. W.H. and R.H. built the frequency comb and measured the frequency noise of the repetition rate. S.B. planned and performed the experiments on the QCL- and $CO_2$-pumped molecular lasers, analyzed the data and wrote the paper.

# Quantum cascade laser-pumped terahertz molecular lasers: frequency noise and phase-locking using a 1560nm frequency comb.

# Supporting Information

**Jean-Francois Lampin[1], Antoine Pagies[1], Giorgio Santarelli[2], Jeffrey Hesler[3], Wolfgang Hänsel[4], Ronald Holzwarth[4] and Stefano Barbieri[1*]**

[1]Institut d'Électronique, de Microélectronique et de Nanotechnologie, CNRS, Université de Lille, 59652 Villeneuve d'Ascq, France
[2] LP2N, IOGS, CNRS, Université de Bordeaux, 33400 Talence, France
[3] Virginia Diodes Inc., Charlottesville, USA
[4] Menlo Systems GmbH, Martinsried, Germany
* e-mail: stefano.barbieri@univ-lille.fr

## S1. Evaluation of the FC frequency noise

In our experiments we used the 1070th and 2500th harmonics of $f_{\mathrm{rep}}$ (~ 1GHz) as LOs to measure the frequency noise of the QCL and $CO_2$-pumped MLs. The estimation of the frequency noise of the harmonics of $f_{\mathrm{rep}}$ was done in two steps. First, we measured the frequency noise of the optical beating ($f_{\mathrm{beat}}$) between the FC and a fiber laser at λ = 1542nm (194.5THz), previously stabilized to a high finesse cavity. Next, we measured the frequency noise of the carrier-envelope offset frequency, $f_{\mathrm{CEO}}$, of the FC using a standard $f$-2$f$ interferometric technique. The results of these measurements are shown in Fig.S1. As shown in the Figure, (i) there is a clear correlation between the two noise spectra, and (ii) the frequency noise of $f_{\mathrm{CEO}}$ considerably exceeds that of $f_{\mathrm{beat}}$.

The frequency of the optical beanote, $f_{\mathrm{beat}}$, is given by the following relation:

$$f_{beat} = \left| f_L - n \times f_{rep} - f_{CEO} \right|, \qquad (1)$$

where $f_L$ is the frequency of the fiber laser and $n = 195 \times 10^3$ ($=f_L/f_{\mathrm{rep}}$). Hence, neglecting the noise contribution of $f_L$, the frequency fluctuation of $f_{\mathrm{rep}}$ is given by:

$$\delta f_{rep} = (\delta f_{beat} + \delta f_{CEO})/n \ . \qquad (2)$$

The frequency noise PSD of $f_{\mathrm{rep}}$ is therefore given by:

$$S_\nu^{f_{rep}}(f) = \left[ S_\nu^{f_{beat}}(f) + S_\nu^{f_{CEO}}(f) + 2\rho \left( S_\nu^{f_{CEO}}(f) \times S_\nu^{f_{beat}}(f) \right)^{1/2} \right] / n^2 \ , \qquad (3)$$

where $-1 \leq \rho \leq 1$ is the correlation coefficient between $f_{\mathrm{beat}}$ and $f_{\mathrm{CEO}}$. In Fig.S2 are reported different frequency noise PSDs of the 2500th harmonic of $f_{\mathrm{rep}}$ derived from Eq.(3). These are obtained by multiplying the spectra of Fig.S1 by the factor $(2.5\mathrm{THz}/195\mathrm{THz})^2$, and using different values of the correlation coefficient, namely: (i) $\rho = +1$ (green line), (ii) $\rho = 0$ (blue line), (iii) $\rho = -1$ (pink line), and (iv) $\rho = -0.7$ (red line). As shown in the Figure, the frequency noise PSD spectrum of the 2.5THz, $CO_2$-pumped methanol laser (black line, same as Fig.4), stays between the $\rho = +1$ (complete correlation - green line) and the $\rho = -1$ (complete anti-correlation - pink line) spectra. This finding provides a strong indication that in the ~3kHz-50kHz range the frequency noise PSD analysis of the $CO_2$-pumped methanol laser is indeed limited by the frequency noise of 2500 x $f_{rep}$. From the agreement between the frequency noise PSD spectrum of the $CO_2$-pumped ML and the red spectrum, obtained for $\rho = -0.7$, we infer a strong degree of correlation between $f_{beat}$ and $f_{CEO}$, (the red spectrum is the one reported in Fig.4). This is in good agreement with the estimated correlation (not shown) between $f_{beat}$ and $f_{CEO}$, obtained from the ratio of the square root of the spectra of Fig.S1, yielding $|\rho| = 0.55$.

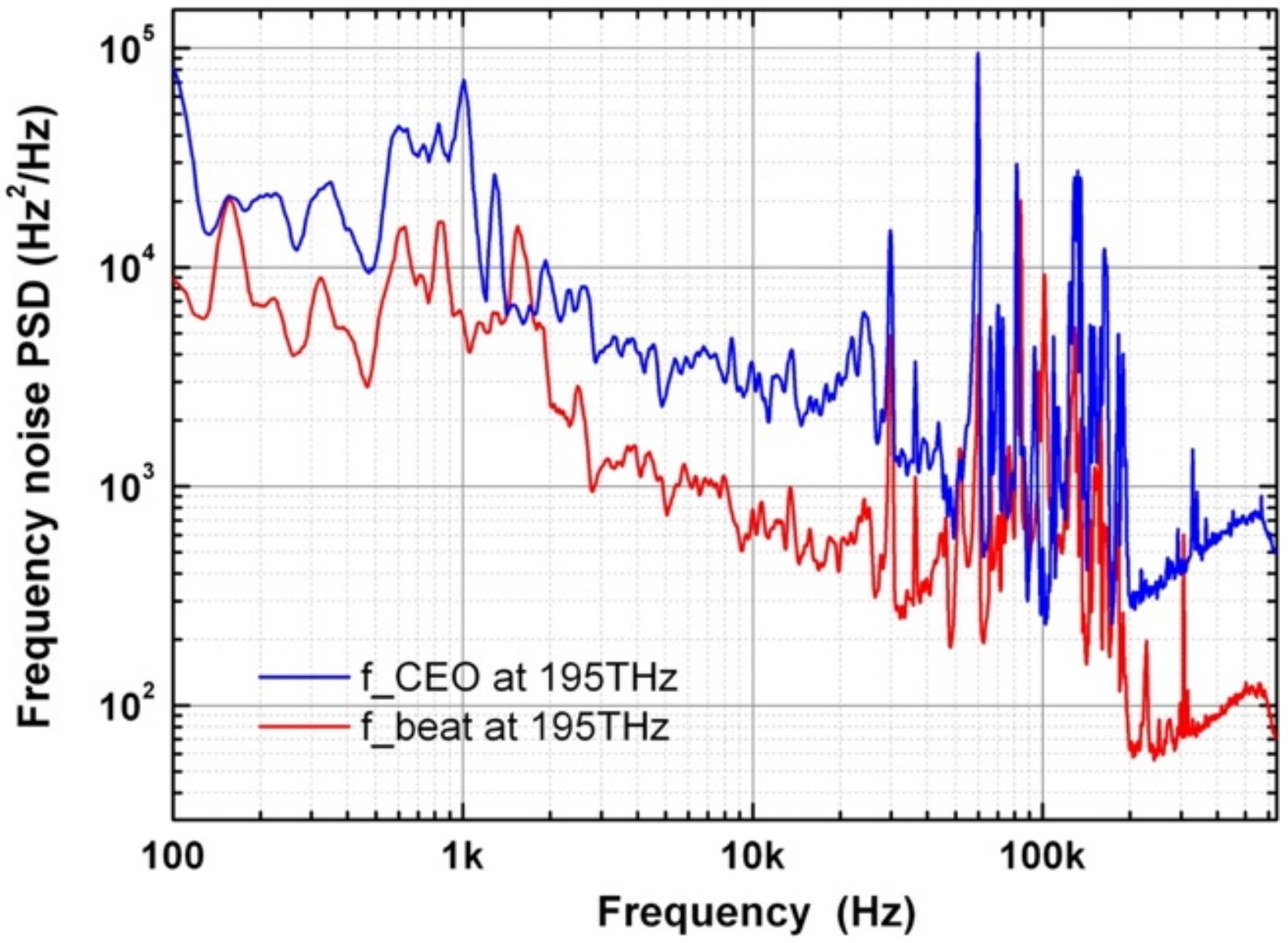

**Fig.S1**. Frequency noise PSD spectrum of $f_{beat}$ and $f_{CEO}$ (at 195THz). The frequency noise PSD spectrum of $f_{beat}$ was obtained by beating the FC with a cavity stabilized fiber laser at λ = 1542nm (194.5THz). The frequency noise PSD spectrum of $f_{CEO}$ was measured using a standard *f*-2*f* interferometric technique.

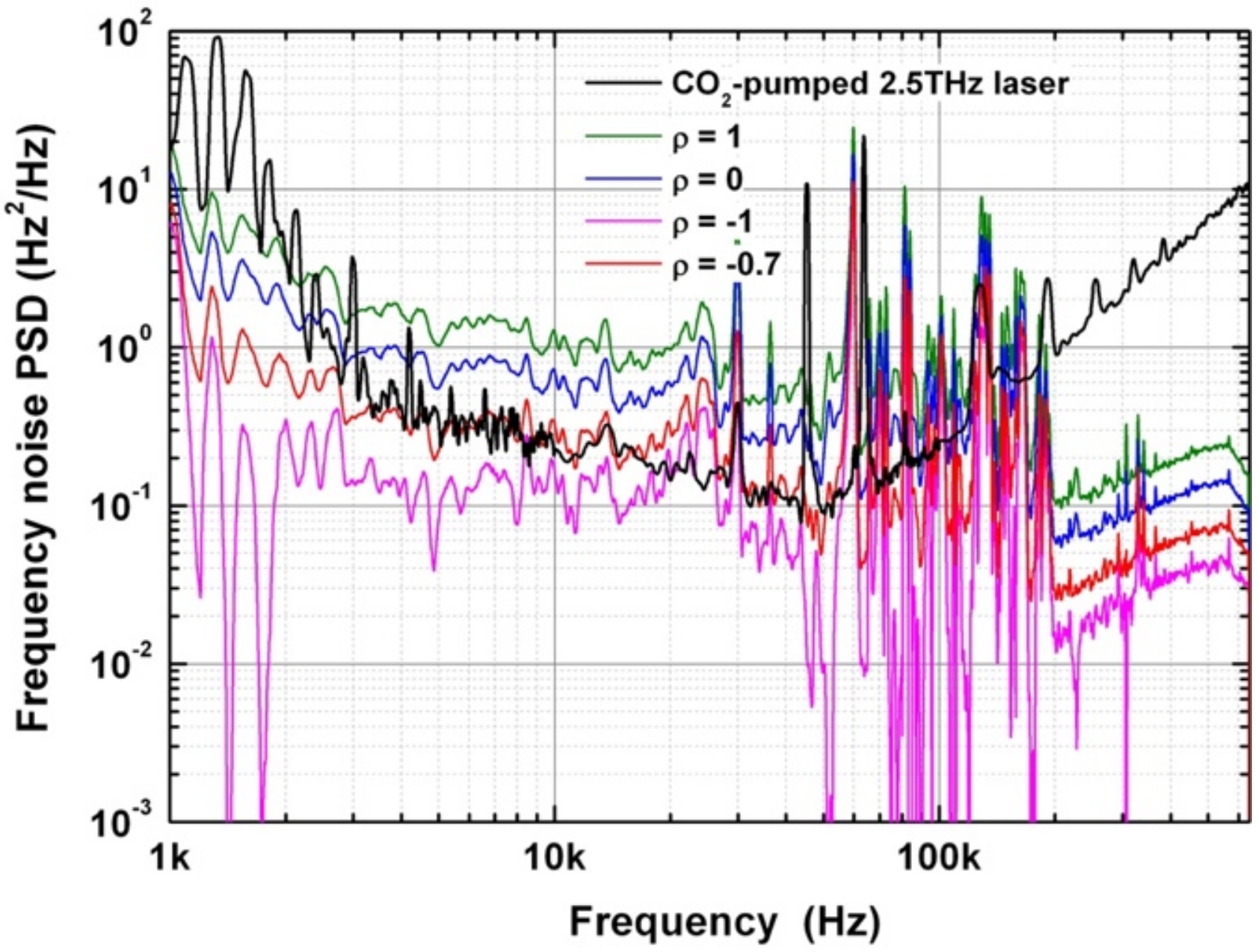

**Fig.S2.** CFrequency noise PSD of the $CO_2$-pumped 2.5THz laser (black line) and frequency noise PSDs of 2500 x $f_{rep}$ obtained from Eq.(4) for different correlation coefficients (see legend). The spectrum corresponding to $\rho = -0.7$ is the one reported in Fig.4.

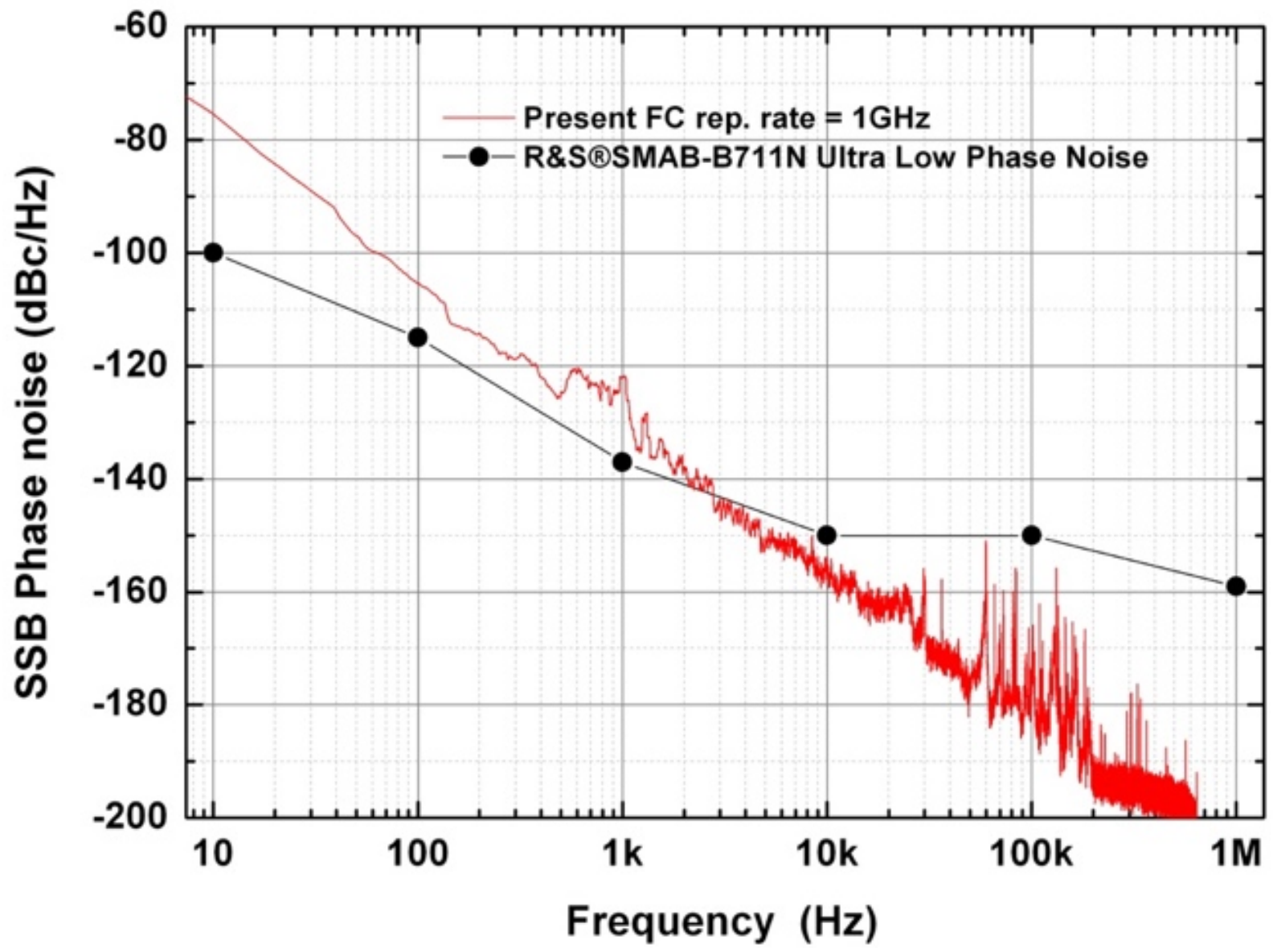

**Fig.S3** Phase noise PSD of $f_{rep}$ = 1GHz (red line) compared to that of a state-of-the-art signal generator (black dots). The phase noise PSD spectrum of $f_{rep}$ is obtained from the measured frequency noise PSD spectra of $f_{beat}$ and $f_{CEO}$ using Eq. (4) with $\rho = -0.7$ . The phase noise PSD of the R&S®SMA100B-B711N was taken from the instrument specifications data [2].

In Fig.S3 we report the single-sideband phase noise PSD of $f_{rep}$ = 1GHz derived from the red spectrum of Fig.S2 ($\rho = -0.7$), with, for comparison, the values of phase noise at 1GHz carrier frequency of a state-of-the-art commercial microwave generator (R&S®SMA100B-B711N Ultra Low Phase Noise) [1]. Above 10kHz from the carrier the phase noise of $f_{rep}$ is clearly lower, showing that the repetition rate of fs-fiber FCs is well suited for THz metrology. It is also worth noting that the obtained frequency noise is between 10 and 20dB lower than that of the fundamentally mode-locked laser ($f_{rep}$ = 250MHz) used in Ref. [2].

.

## S2. Estimation of the pump transition saturation

To estimate the saturation intensity of the saQ(3,3) transition we have used the following equation [3]:

$$I_s = \frac{8\pi h\nu^3}{c^2}\frac{1}{4T_1T_2\times A_{ba}g_b}\left[\frac{1}{9A_i^X g_\pm}\right], \qquad (4)$$

with $T_1$ and $T_2$ the population and dipole relaxation rates respectively, $A_i^X \sim 0.03$, an angular factor that we averaged over all possible hyperfine transitions considering a linearly polarized beam [48], $A_{ba}$ the Einstein coefficient, $g_b$ the degeneracy of the upper level and $g_\pm$ a geometrical factor equal to 1⁄2 for a Gaussian beam. At a pressure of ~1Pa, $T_2 \sim 1\mu s$ [49] and $T_1 \sim 500ns$, yielding $I_s \sim 2mW/cm^2$, which, multiplied by the section of the waveguide, corresponds to ~0.5mW of saturation power.

## S3. Experimental setup for the measurement of the frequency noise PSD of the $CO_2$-pumped ML

As shown in Fig.S4, the $CO_2$-pumped ML and the FC are collinearly focused on an electro-optic detection unit, consisting of a 2mm thick ZnTe crystal, followed by a quarter- and a half-wave plates, a Wollaston prism, and a home-made balanced detection [2, 4]. The latter is based on two fiber-coupled InGaAs photodiodes and has a RF bandwidth of 500MHz. For the experiment we used approximately 50mW of power (λ= 780nm) impinging on each photodiode, providing a shot-noise limited noise-floor. The $CO_2$-pumped methanol laser, generates approximately 50mW of output power. Due to water absorption along the optical path and beam divergence we measured ~20mW of THz power on the ZnTe crystal.

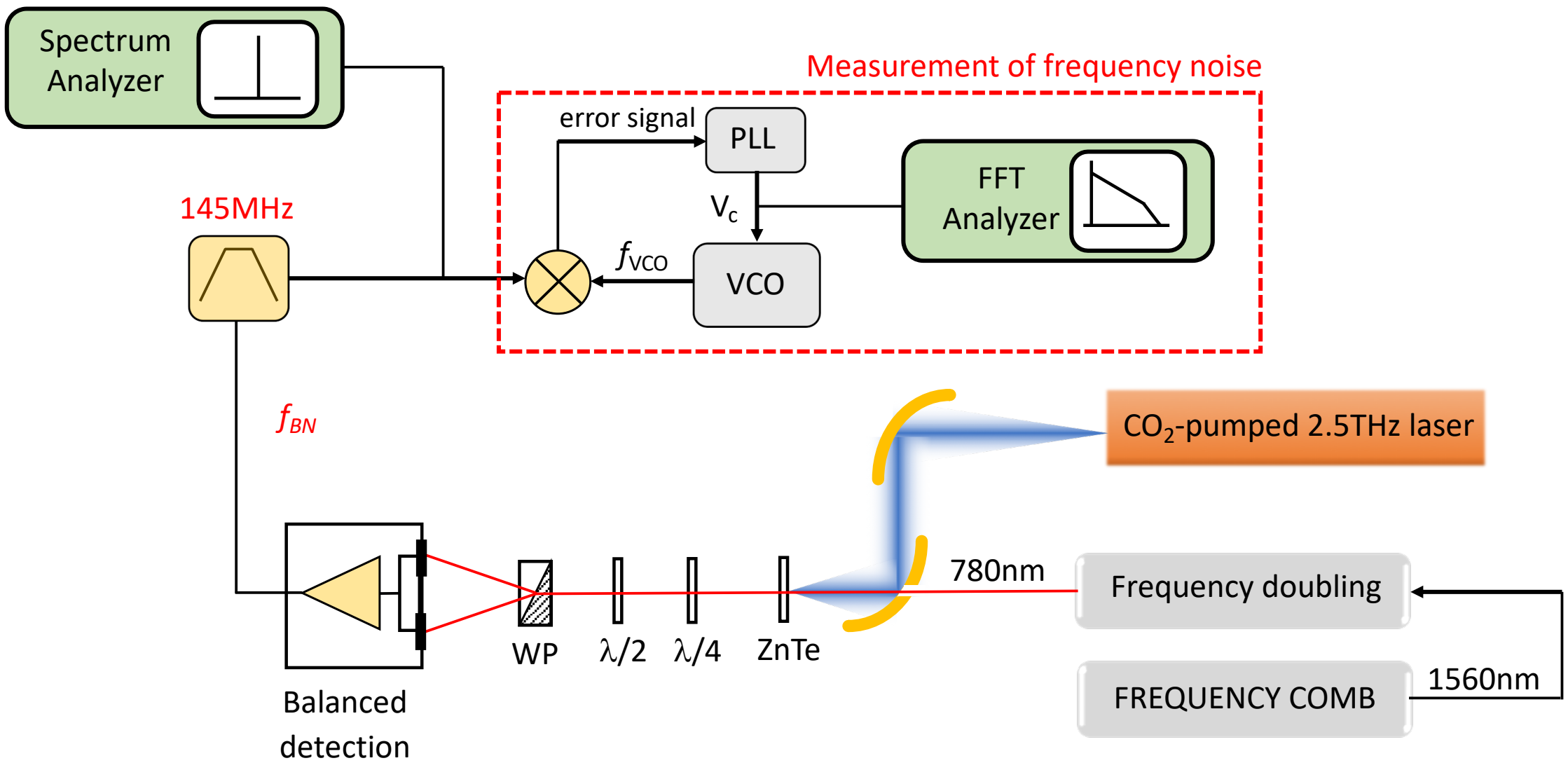

**Fig.S4**. Experimental setup for the frequency noise measurement of the $CO_2$-pumped ML (see text). The beatnote frequency $f_{BN}$ is filtered and demodulated using the same technique, based on a VCO, used for the QCL-pumped ML and described in the main text.

At the output of the balanced detection we obtain a beatnote $f_{BN}= f_{THz} - n \times f_{rep} = 130MHz$ where $f_{THz}= 2.5THz$ is the frequency of the FIR laser and $N = \text{Int}\,(f_{THz}/f_{rep}) \sim 2500$ [4]. The frequency noise PSD of the ML is derived from $f_{BN}$ with the setup displayed in the red-frame of Fig.S4, based on the same demodulation technique used for the QCL-pumped ML described in the main text.

## S4. Comparison with THz QCLs and state-of-the-art multiplied oscillators

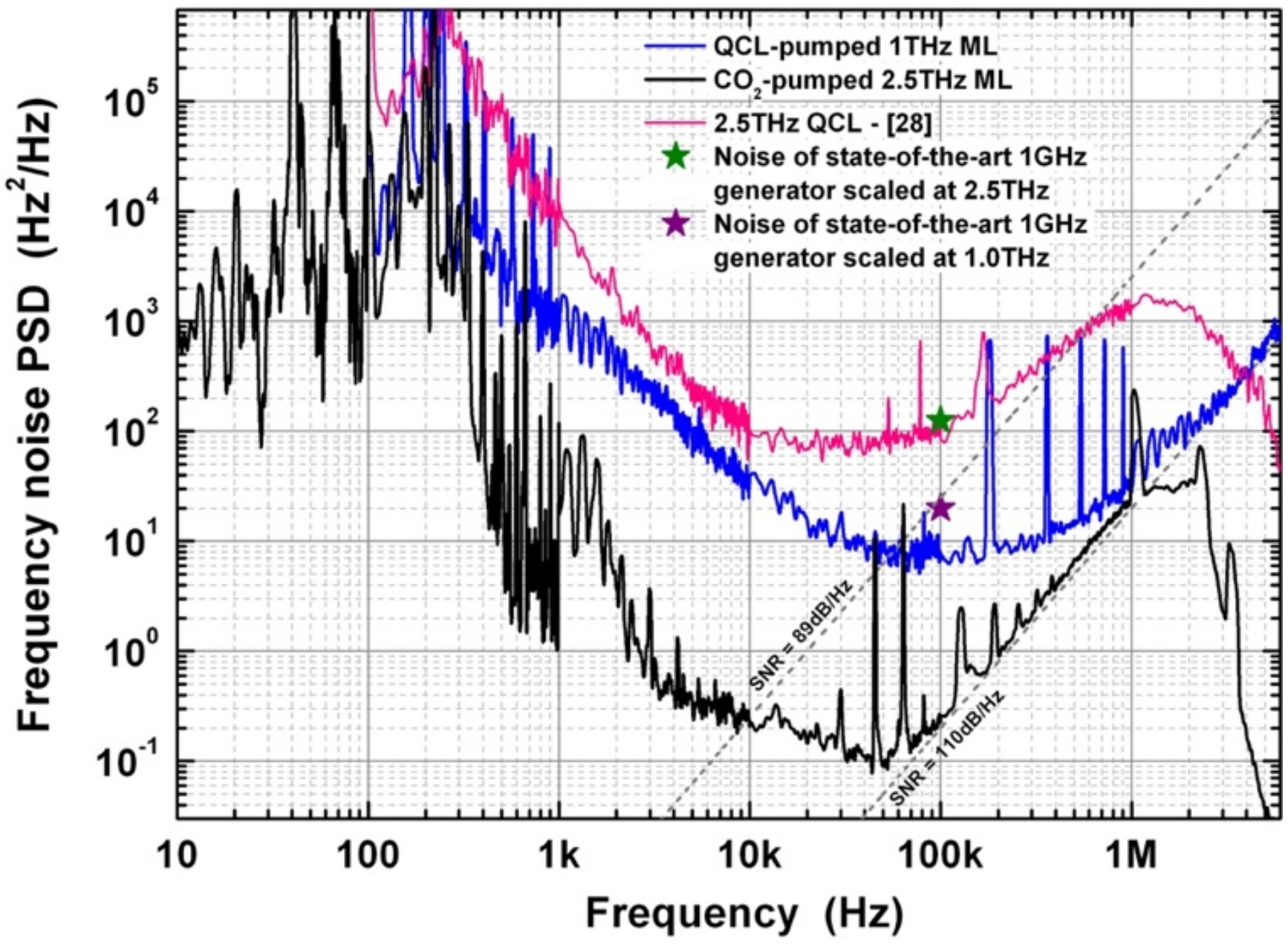

**Fig.S5**. Frequency noise PSD spectra of (i) the QCL-pumped, 1.073THz ML (blue), (ii) the $CO_2$-pumped 2.5THz ML (black), and (iii) the 2.5THz QCL of Ref. [28], operating at T=20K (purple). The red and green stars are the values of the frequency noise of the R&S®SMA100B-B711N at 1GHz (see Fig.S3), extrapolated at 2.5THz and 1THz (100kHz from carrier).

In Fig. S5 we report a comparison between the frequency noise PSDs of the present QCL-pumped ML laser emitting at 1THz (blue) and that of a 2.5THz QCL measured in another work (purple) [28]. We note that for the latter, the plateau in the 10-100kHz range corresponds to the Shawlow-Townes limit of the source.